\documentstyle[preprint,aps,prd]{revtex}

\begin{document}

\title {Einstein-Langevin Equations from Running Coupling Constants}
    
\author{Fernando C.\ Lombardo \footnote{Electronic address: 
lombardo@df.uba.ar}}

\address{{\it
Departamento de F\'\i sica, Facultad de Ciencias Exactas y Naturales\\ 
Universidad de Buenos Aires- Ciudad Universitaria, Pabell\' on I\\ 
1428 Buenos Aires, Argentina}}

\author{Francisco D.\ Mazzitelli 
\footnote{Electronic address: fmazzi@df.uba.ar}}

\address{{\it
Departamento de F\'\i sica and IAFE\\
 Facultad de Ciencias Exactas y Naturales\\ 
Universidad de Buenos Aires- Ciudad Universitaria, Pabell\' on I\\ 
1428 Buenos Aires, Argentina\\}}

\maketitle

\begin{abstract}
The Einstein-Langevin equations  take into account 
the
backreaction of quantum matter fields on the background
geometry. 
We present a derivation of these equations to lowest 
order in a covariant expansion in powers of the curvature. 
For massless fields, the equations are completely determined 
by the running coupling constants of the theory.
\end{abstract}
\newpage

The study of the backreaction of quantum matter fields on the
spacetime geometry is a very interesting subject. It is relevant
to understand the fate of black hole evaporation \cite{review}, 
for the
analysis of the eventual smearing of the classical cosmological
singularities due to quantum effects \cite {hartlehu},  
for the study of
the theoretical possibility of creating  closed timelike curves
(time machines) \cite{wald}, etc.

The usual approach to this problem is based on the use
of the Semiclassical Einstein Equations (SEEs) \cite{birrel}
\begin{equation}
{1\over{8\pi G}}\left[R_{\mu\nu}-\frac{1}{2}Rg_{\mu\nu}\right]- 
\alpha H^{(1)}_{\mu\nu} - \beta H^{(2)}_{\mu\nu}= T_{\mu\nu}^{clas}+ 
<T_{\mu\nu}>,
\label{see}
\end{equation}
where the effect of quantum matter fields is taken into account
by including as a source the quantum mean value of the 
energy-momentum tensor. The terms proportional to
\begin{equation} 
H^{(1)}_{\mu\nu} = \left[4 R^{;\mu\nu}-4g^{\mu\nu}\Box R\right]+ 
O(R^2),\end{equation} and 
\begin{equation} H^{(2)}_{\mu\nu}=\left[4R^{\mu\alpha;\nu}{}_{;\alpha}-2\Box 
R^{\mu\nu}- g^{\mu\nu}\Box R\right]+ O(R^2),\end{equation}
come from terms quadratic in the curvature  in the gravitational action,
which are
needed to renormalize the theory.
The SEEs take into account dissipative
effects
of the quantum fields on the metric,
and can be derived
from the {\it real part} of the Closed Time Path (CTP) effective
action \cite{varios}.

It is clear that these equations cannot provide a full description
of the problem, at least when 
the state of the quantum matter fields is such that the 
energy momentum tensor has important fluctuations around
its mean value \cite{ford}.
These fluctuations can be taken into account by  
including an additional stochastic term \cite{calhu,camver0,camver}
on the right hand side of Eq. \ref{see}. This noise-term comes
from the imaginary part of the CTP effective action.
The SEE thus become ``Einstein-Langevin Equations" (ELEs), that include
both the dissipative and diffusive effects of the quantum matter
on the geometry of spacetime \cite{calhu}. 

In  previous works, the ELEs were 
derived for arbitrary small metric perturbations conformally coupled
to a massless quantum scalar field in a spatially flat background 
\cite{camver}, and, in a cosmological setting, for a massive field 
in a spatially flat Friedmann-Robertson-Walker universe \cite{humatacz}, 
and in a Bianchi type-I spacetime\cite{husinha}.

In this paper we will present a derivation of the ELEs 
for a quantum scalar field,
to lowest order in a
covariant expansion in powers of the curvature.
We will see that, when the scalar field is massless, the ELEs are
fully determined
by the running couplings of the theory. For massive fields,
some additional work is needed in order to obtain the 
corresponding dissipation and noise kernels.

Our strategy will be as follows. We will first compute the Euclidean
effective action up to second order in the curvature.
With an adequate replacement of Euclidean propagators by 
propagators along a closed time path, we will obtain the CTP effective
action. The dissipation and noise kernels in the
CTP action will allow as to obtain
the ELEs.

Consider a massive scalar field on a classical, Euclidean curved 
background.
The classical action is given by
\begin{equation}
S=S_{grav} + S_{matter},
\label{sclas}
\end{equation}
where
\begin{equation}
S_{grav}=-\int d^4x\sqrt g\left [{1\over 16\pi G_0}(R-2\Lambda_0)
+\alpha_0 R^2 +\beta_0 R_{\mu\nu}R^{\mu\nu}\right ],\label{bare}
\end{equation}
and
\begin{equation}
S_{matter}={1\over 2}\int d^4x\sqrt g [\partial_{\mu}\phi\partial^{\mu}\phi
+m^2\phi^2+\xi R\phi^2].
\end{equation}
Here $\xi$ is the coupling to the curvature. $G_0$, $\Lambda_0$, 
and the dimensionless 
constants $\alpha_0$ and $\beta_0$ are {\it bare} constants.

The effective action for this theory is a complicated, non local
object. It is defined by integrating out the quantum scalar field, that is
\begin{equation} e^{-S_{eff}} = N \int {\cal D}\phi e^{-S[g_{\mu\nu}, 
\phi]},\end{equation} where $N$ is a normalization constant. It is in general 
not possible to find a closed form for it. On general grounds, we expect it to be 
of the form \cite{vilkoqt} \begin{eqnarray} S_{eff} &=& -\int d^4x\sqrt g\left 
[{1\over 16\pi G}R +\alpha R^2 +\beta R_{\mu\nu}R^{\mu\nu}\right ]\nonumber\\ 
&+&\frac{1}{32\pi^2} \int d^4x\sqrt g\left[F_0 R +  RF_1(\Box)R+R_{\mu\nu} 
F_2(\Box)R^{\mu\nu}+...\right], \label{seff} \end{eqnarray} where the ellipsis 
denote terms cubic in the curvature. For simplicity, in the above equation and 
in what follows we will  omit the cosmological constant term. Note that the 
bare constants in Eq. \ref{bare} have been replaced by dressed couplings in 
Eq. \ref{seff}. 

Up to this order, all the information about the effect of the quantum
field is encoded in the constant $F_0$ and in the form 
factors $F_1$ and $F_2$.
The form factors are, in general, non-local two point functions constructed
with the d'Alambertian and the parameters
$\xi$ and $m^2$.  $F_0$, $F_1$, and $F_2$ also depend on an energy
scale $\mu$, introduced by the regularization method. 

The dressed coupling constants depend on the energy scale
$\mu$ according to the renormalization  group equations
\begin{eqnarray}
\mu{dG\over d\mu}&=&
 \frac{G^2 m^2}{\pi} \left(\xi -  \frac{1}{6}  \right),\label{rge1} \\
\mu{d\alpha\over d\mu}&=&
 -\frac{1}{32 \pi^2}
 \left[
 \left( \frac{1}{6} - \xi \right)^2 -\frac{1}{90}
 \right], \label{rge2}\\
\mu{d\beta\over d\mu}&=&
 -\frac{1}{960 \pi^2}. \label{rge3}
\end{eqnarray}
The dependence of $F_0$, $F_1$ and $F_2$ on $\mu$ is such that
the full equation is $\mu$-independent. For example, 
from Eqs. \ref{seff} and \ref{rge1}, 
we see that $F_0=m^2\ln\left({m^2\over\mu^2}\right)(\xi-{1\over 6})+const$.

When the scalar field is massless, this information is
enough to fix completely the form factors. Indeed, as the
$F_i,\,\, i=1,2$ are  dimensionless two point functions, 
by simple dimensional analysis we obtain $F_i(\Box,\mu^2,\xi)=
F_i({\Box\over\mu^2},\xi)$. Inserting this into
Eq. \ref{seff}, using Eqs. \ref{rge2} and \ref{rge3}, and the fact  that $S_{eff}$ 
must be independent of
$\mu$, we obtain
\begin{eqnarray}
F_1(\Box)&=&{1\over 2}\left[(\xi-{1\over 6})^2-{1\over 90}\right ]
\ln \left [{-\Box
\over\mu^2}\right] + const,\nonumber\\                   
F_2(\Box)&=&{1\over 60} 
\ln \left [{-\Box
\over\mu^2}\right] + const.
\label{m=0}
\end{eqnarray}

The final result for the effective action has a clear interpretation:
it is just the classical action in which the coupling
constants  $\alpha$ and $\beta$ have been replaced by non-local 
two point functions 
that take into 
account their running  in {\it configuration
space}.

For a massive field, the situation is more complex because there
is an additional dimensional parameter. The form factors also
depend on ${m^2\over\mu^2}$ and the $\mu$-independence of the
effective action is not enough to fix the form of them. Fortunately,
they have already been computed in the literature \cite{barvi,avra} 
\begin{equation}
F_i(\Box)=\int_0^1 d\gamma \chi_i(\xi, \gamma) \ln \left 
[{m^2-{1\over 4}(1-\gamma^2)\Box
\over\mu^2}\right]\label{fgrav},\end{equation}
where 
\begin{eqnarray}
\chi_1(\xi, \gamma)&=&{1\over 2} \left [ \xi^2-{1\over 2}\xi(1-\gamma^2)
+{1\over 48}(3-6 \gamma^2-\gamma^4)\right ],\nonumber\\
\chi_2(\xi, \gamma)&=&{1\over 12}\gamma^4.
\label{chi}
\end{eqnarray}
These equations can be obtained through a covariant
perturbation expansion \cite{barvi}, or by a resummation
of the Schwinger DeWitt expansion \cite{avra}. 
Of course these form factors coincide with our previous Eq. \ref{m=0}
in the massless case.

In order to clarify the meaning of the two point functions
appearing in Eqs. \ref{m=0} and \ref{fgrav}, it is
useful to introduce the following integral representation \cite{dd}
\begin{equation}
\ln\left[{m^2-{1\over 4}(1-\gamma^2)\Box\over\mu^2}\right]
=\left[\ln
{(1-\gamma^2)\over 4}+\int_0^{\infty}dz\left({1\over z+\mu^2}- 
G_E^{(z)}\right)\right],\label{rep}
\end{equation}
so the logarithm of the d'Alambertian is written in 
terms of the massive Euclidean
propagator $G_E^{(z)} = (z+{4 m^2\over (1-\gamma^2)}-\Box )^{-1}$.
This representation will also be useful to construct the CTP
version of the effective action.

Replacing the Euclidean propagator by the Feynman one 
in the integral representation Eq. \ref{rep}, one
obtains the usual {\it in-out} effective action. As is very
well known, the effective equations derived from this action are neither real
nor causal because they are equations for {\it in-out} matrix
elements and not for mean values.

The solution to this problem is also well known. Using the CTP formalism
one can construct an {\it in-in} effective action that produces
real and causal field equations for {\it in-in} expectation values
\cite{varios}. The 
effective action can be written as
\begin{equation} e^{i S_{eff}[g^+,g^-]} = N e^{i(S_{grav}[g^+]-S_{grav}[g^-])} 
\int {\cal D} \phi^+{\cal D} \phi^- e^{i(S_{matter}[g^+,\phi^+]- S_{matter}[g^-,\phi^-])}, 
\label{ctpeff}\end{equation} and the field equations are obtained taking the 
variation of this action with respect to the $g_{\mu\nu}^+$ metric, and then 
setting $g_{\mu\nu}^+=g_{\mu\nu}^-$. 

In an alternative, and more concise notation, we can write this effective
action as
\cite{mottola}
\begin{equation} e^{i S_{eff}^{{\cal C}}[g]} = N e^{i S_{grav}^{{\cal C}}[g]} 
\int {\cal D} \phi e^{i S_{matter}^{{\cal 
C}}[g,\phi]},\label{neweff}\end{equation} where we have introduced the CTP 
complex temporal path ${\cal C}$, going from minus to plus infinity $\cal C_+$ 
and backwards $\cal C_-$, with a decreasing (infinitesimal) imaginary part. 
Time integration over the contour ${\cal C}$ is defined by $\int_{{\cal C}} dt 
=\int _{{\cal C_+}} dt -\int_{{\cal C_-}} dt$. The field $\phi$  appearing in 
Eq. \ref{neweff} is related to those in Eq. \ref{ctpeff} by $\phi(t,\vec x) = 
\phi_{\pm}(t,\vec x)$ if $t \in {\cal C}_{\pm}$. The same applies to 
$g_{\mu\nu}$. 

This equation is 
useful because it has the 
structure of the usual {\it in-out} or the Euclidean effective action. Feynman
rules are therefore the ordinary ones, replacing Euclidean propagator by
\begin{eqnarray} G(x,y) = \left\{\begin{array}{ll} 
G_F(x,y)=i \langle 0, in\vert T \phi (x) \phi(y)\vert 0, in\rangle,& ~t, t' ~ 
\mbox{both on} ~{\cal C}_+ \\ G_D(x,y)=-i \langle 0, in\vert {\tilde T}\phi (x) 
\phi(y)\vert 0, in\rangle , & ~t, t' ~ \mbox{both on} ~{\cal C}_-\\ G_+(x,y)=-
i \langle 0, in\vert \phi (x) \phi(y)\vert 0, in\rangle,    &~t ~\mbox{on}~ 
{\cal C}_-, t'  ~\mbox{on} ~{\cal C}_+\\ G_-(x,y)=i \langle 0, in\vert \phi (y) 
\phi(x)\vert 0, in\rangle, & ~ t  ~\mbox{on} ~ {\cal C}_+, t'~  \mbox{on}~ 
{\cal C}_-\end{array}\right. \label{prop} \end{eqnarray} 

Introducing Riemann normal coordinates, we can write, up to lowest order in 
the curvature 
 
\begin{equation} G_F(x,y)= \int {d^4p\over{(2 \pi)^4}} {e^{ip(x-
y)}\over{p^2+m^2-i\epsilon}}= G_D^*(x,y),\end{equation} 

\begin{equation} 
G_{\pm}(x,y)=\mp \int {d^4p\over{(2 \pi)^4}} e^{ip(x-y)}2 \pi i 
\delta(p^2-m^2)\theta(\pm p^0).
\label{gpm}
\end{equation}

All of the preceding formulation of the effective action is valid for 
any field theory. In our particular case, we must replace the 
Euclidean propagator $G_E^{(z)}$ in Eq. \ref{rep} by the propagator 
$G(x,y)$ of Eq. \ref{prop} with a mass given by  ${4m^2\over{1-\gamma^2}}
+z$. After integration in $z$ we obtain
\begin{eqnarray}
\ln\left[{{4 m^2\over{(1-\gamma^2)}}- \Box\over\mu^2}\right]_{CTP}
= \left\{\begin{array}{ll}
 \int {d^4p\over{(2 \pi)^4}}e^{ip(x-y)}  \ln\left({(1-\gamma^2)
(p^2-i\epsilon)+4m^2\over{\mu^2}}\right) & ~t, t' ~ \mbox{both on} ~{\cal C}_+
\\
\int {d^4p\over{(2 \pi)^4}}e^{ip(x-y)}  \ln\left({(1-\gamma^2)
(p^2+i\epsilon)+4m^2\over{\mu^2}}\right) & ~t, t' ~ 
\mbox{both on} ~{\cal C}_-\\
\int {d^4p\over{(2 \pi)^4}}e^{ip(x-y)}2 \pi i \theta (p^0)\theta 
\left(-p^2 -{4m^2\over{1-\gamma^2}}\right)    &~t ~\mbox{on}~ {\cal C}_-, t'  
~\mbox{on} ~{\cal C}_+\\
-\int {d^4p\over{(2 \pi)^4}}e^{ip(x-y)}2 \pi i \theta (-p^0)\theta 
\left(-p^2 -{4m^2\over{1-\gamma^2}}\right) & ~ t  ~\mbox{on} ~ {\cal C}_+, t'~  
\mbox{on}~ {\cal C}_-\end{array}\right.\end{eqnarray}

With the expression for the CTP logarithm of the d'Alambertian we can calculate
explicitly the CTP effective action. Using the previous 
notation with $g_{\mu\nu}^+$ and $g_{\mu\nu}^-$ the CTP effective 
action reads
 
\begin{eqnarray}
S_{eff}&&[g^+,g^-] = S^r_{grav}[g^+] - S^r_{grav}[g^-] \nonumber \\
&&+{i\over{8 \pi^2}}\int d^4x \int d^4y \Delta(x) \Delta(y)N_1(x,y)
- {1\over{8 \pi^2}}\int d^4x \int d^4y \Delta(x)\Sigma(y)D_1(x,y)\nonumber\\
&&+{i\over{8 \pi^2}}\int d^4x\int d^4y\Delta_{\mu\nu}(x)
\Delta^{\mu\nu}(y)N_2(x,y)
-{1\over{8 \pi^2}}\int d^4x\int d^4y \Delta_{\mu\nu}(x)\Sigma^{\mu\nu}(y)
D_2(x,y),
\label{seffctp}
\end{eqnarray}
where $\Delta = {{R^+ - R^-}\over{2}}$, $\Sigma = {{R^+ + R^-}\over{2}}$, 
$\Delta_{\mu\nu} = {{R^+_{\mu\nu} - R^-_{\mu\nu}}\over{2}}$, 
$\Sigma_{\mu\nu} = {{R^+_{\mu\nu}
 + R^-_{\mu\nu}}\over{2}}$. The classical gravitational
action $S^r_{grav}$ contains the dressed, $\mu$-dependent coupling 
constants and we absorbed $F_0$ into the gravitational constant $G$.

The real and imaginary parts of $S_{eff}$ can be associated with 
the dissipation and noise, 
respectively.  The dissipation 
$D_i$ and 
noise $N_i$ kernels are given by

\begin{equation} D_i(x,y) = \int_0^1 d\gamma \chi_i(\xi, \gamma)\int 
{d^4p\over{(2 \pi)^4}} \cos[p (x-y)]\ln\left|{{(1-\gamma^2)p^2 + 
4 m^2}\over{\mu^2}}\right|,\end{equation}

\begin{equation} N_i(x,y) = \int_0^1 d\gamma \chi_i(\xi, \gamma)\int 
{d^4p\over{(2 \pi)^4}} \cos[p(x-y)] \theta\left(- p^2 - {4 m^2\over{1-
\gamma^2}}\right). \end{equation} 

It is important to note that the imaginary part of this effective action must 
be positive definite. To make this point  explicit, one can write the imaginary 
part in terms of the Weyl tensor $C_{\mu\nu\alpha\beta}$ and the scalar 
curvature $R$ by means of the following relation: 
$C_{\mu\nu\alpha\beta}C^{\mu\nu\alpha\beta}=2R_{\mu\nu}R^{\mu\nu} - 2/3 R^2$. 
It is not difficult to show that the scalar and tensor contributions to the 
imaginary part of the effective action are both positive. 

One can regard the imaginary part of the closed-time-path-effective-action 
(CTPEA) as 
coming from two classical stochastic sources $\eta(x)$ and 
$\eta^{\mu\nu\alpha\beta}(x)$, where the last tensor has the symmetries of the 
Weyl tensor. In fact, as usually done in statistical physics, we can write the 
imaginary part of the CTPEA as 

\begin{eqnarray} \int {\cal D}\eta(x)&& \int {\cal 
D}\eta^{\mu\nu\alpha\beta}(x) P[\eta, 
\eta^{\mu\nu\alpha\beta}]\exp\left(i\left\{ \Delta(x) \eta(x) + 
\Delta_{\mu\nu\alpha\beta} \eta^{\mu\nu\alpha\beta}\right\}\right) \nonumber 
\\ &&= \exp\left\{-\int d^4x\int d^4y \left[\Delta(x) {\tilde N}(x-y) 
\Delta(y) + \Delta_{\mu\nu\alpha\beta}(x) N_2(x-y) 
\Delta^{\mu\nu\alpha\beta}(y)\right]\right\}, \end{eqnarray} where ${\tilde 
N}(x,y) = N_1(x,y)+ 1/3 N_2(x,y)$, and $\Delta_{\mu\nu\alpha\beta} = 
{C^+_{\mu\nu\alpha\beta} - C^-_{\mu\nu\alpha\beta}\over{2}}$. $P[\eta, 
\eta^{\mu\nu\alpha\beta}]$ is a Gaussian functional probability distribution 
given by \begin{eqnarray} P[\eta, \eta^{\mu\nu\alpha\beta}] = A\,\, && 
\exp\left\{-{1\over{2}} \int d^4x\int d^4y \eta(x) \left[ {\tilde N}(x, 
y)\right]^{-1} \eta(y)\right\} \nonumber \\ &&\times\exp\left\{-
{1\over{2}}\int d^4x\int d^4y \eta_{\mu\nu\alpha\beta}(x) \left[ N_2(x, 
y)\right]^{-1} \eta^{\mu\nu\alpha\beta}(y)\right\},\label{noise}\end{eqnarray} 
with $A$ a normalization factor. 

Therefore, the CTPEA can be written as
\begin{equation} \exp\{iS_{eff}\}= \int {\cal D}\eta {\cal 
D}\eta_{\mu\nu\alpha\beta} P[\eta ,\eta_{\mu\nu\alpha\beta}] \exp\left\{i 
A_{eff}[\Delta, \Delta_{\mu\nu\alpha\beta}, \Sigma, \Sigma_{\mu\nu}, \eta, 
\eta_{\mu\nu\alpha\beta}]\right\},\end{equation} where 
\begin{equation} A_{eff} = Re 
S_{eff} + \int d^4x [\Delta(x) \eta(x) + \Delta_{\mu\nu\alpha\beta}(x) 
\eta^{\mu\nu\alpha\beta}].\end{equation} 

The field equations $\left.{\delta A_{eff}\over{\delta g^+_{\mu\nu}}} 
\right|_{g^+_{\mu\nu} = g^-_{\mu\nu}} = 0$, the Einstein-Langevin equations,
are

\begin{eqnarray}
&&{1\over 8 \pi G}\left( R^{\mu\nu}- 
{1\over{2}} g^{\mu\nu} R \right)
- \tilde{\alpha}H^{(1)}_{\mu\nu}
- \tilde{\beta}H^{(2)}_{\mu\nu}\nonumber \\
&&= -{1\over{32 \pi^2}}\int d^4y   D_1(x,y) H_{\mu\nu}^{(1)}(y)
-{1\over{32 \pi^2}}\int d^4y  D_2(x,y)H_{\mu\nu}^{(2)}(y)\nonumber \\ 
&&+ g^{\mu\nu}\Box \eta - \eta^{;\mu\nu}
+2 \eta^{\mu\alpha\nu\beta}{}{}{}{}_{;\alpha\beta}~~,
\label{ele}
\end{eqnarray}
where $\tilde{\alpha}$ and $\tilde{\beta}$ differ from 
$\alpha$ and $\beta$ by $\xi$-dependent finite constants. 
The Eq. \ref{ele} is our main result.
The r.h.s. consists of the mean value of the 
energy-momentum tensor for the scalar field plus a stochastic correction
characterized by the two point correlation functions
\begin{eqnarray}
<\eta (x)\eta (y)>&=& {\tilde N}(x,y)\nonumber\\
<\eta_{\mu\nu\alpha\beta} (x)\eta_{\rho\sigma\lambda\tau} (y)>&=& 
T_{\mu\nu\alpha\beta\rho\sigma\lambda\tau} N_2(x,y) ,\label{corrfunct} 
\end{eqnarray} where the tensor $T_{\mu\nu\alpha\beta\rho\sigma\lambda\tau}$ 
is a linear combination of four-metric products in such a way that the r.h.s 
of Eq. \ref{corrfunct} keeps the Weyl's symmetries (it is explicitly given in 
the Appendix of Ref. \cite{camver}). The scalar-noise kernel is given by 
\begin{equation} {\tilde N}(x,y)={1\over{2}}\int_0^1 d\gamma \left[\left(\xi-
{(1-\gamma^2)\over{4}}\right)^2-{\gamma^4\over{36}}\right]\int {d^4p\over{(2 
\pi)^4}} cos[p(x-y)]\theta\left(-p^2-{4m^2\over{1-
\gamma^2}}\right).\end{equation} In the massless case ${\tilde N}$ is proportional to 
$(\xi - 1/6)^2$, and vanishes for conformal coupling. Therefore this term is 
present when the quantum fields are massive and/or when the coupling is not 
conformal. This is to be expected, since the imaginary part of the CTPEA is 
directly associated to gravitational particle creation. For massless, 
conformally coupled quantum fields, particle creation takes place only when 
spacetime is not conformally flat. Therefore in this case the only 
contribution to the imaginary part of the CTPEA is proportional to the square 
of the Weyl tensor. When the fields are massive and/or non-conformally 
coupled, particle creation takes place even when the Weyl tensor vanishes. 
This is why an additional contribution proportional to $R^2$ appears in the 
imaginary part of the effective action. 

From Eq. \ref{ele} we can define the effective energy-momentum tensor 
\begin{equation} 
T_{\mu\nu}^{eff}=<T_{\mu\nu}>+T_{\mu\nu}^{stoch}= <T_{\mu\nu}>+ 
g^{\mu\nu}\Box \eta - \eta^{;\mu\nu}
+2 \eta^{\mu\alpha\nu\beta}{}{}{}{}_{;\alpha\beta}~~,\end{equation} where $<T_{\mu\nu}>$ is the quantum expectation value of the energy-
momentum tensor of the quantum field and $T_{\mu\nu}^{stoch}$ 
is the contribution of 
the stochastic force, which in turn has contributions from the scalar and 
tensor noises. In the massless-conformal case the scalar-noise kernel 
vanishes, and $(T_\mu{}^\mu)^{stoch}= 0$, because the noise-source 
$\eta^{\mu\nu\alpha\beta}$ has vanishing trace. This means that there is no 
stochastic correction to the trace anomaly \cite{camver}. 

To summarize, we have obtained the ELEs using a covariant expansion
in powers of the curvature. Our results are valid for quantum
scalar fields with arbitrary mass and coupling to the curvature
$\xi$, thus
generalizing previous results \cite{camver}.
In the massless case, still for arbitrary $\xi$, we have shown that
it is possible to obtain the noise and dissipation kernels using 
only dimensional analysis and the running of the coupling constants. 

We are currently computing the CTPEA and the ELEs in models of dilaton 
gravity
in two dimensions, in order to discuss the relevance of the 
stochastic effects in the quantum to classical transition
of the background metric.

This work was supported by Universidad
de Buenos Aires, CONICET and Fundaci\' on Antorchas. We are grateful to Diego
A. R. Dalvit for many useful discussions.

\end{document}